\documentclass[preprint,showpacs,preprintnumbers,amsmath,amssymb]{revtex4-1}
\usepackage{latexsym}
\usepackage{epsfig}
\usepackage{float}
\usepackage{lipsum}
\usepackage{graphicx,times}
\usepackage{amssymb,amsmath}
\usepackage{color}
\usepackage{dcolumn}
\usepackage{epstopdf}
\usepackage{pgfplots}
\usepackage[colorlinks=true, urlcolor=blue, linkcolor=red, citecolor=blue]{hyperref}
\usepackage{booktabs}
\newcommand{\be}{\begin{equation}}
\newcommand{\ee}{\end{equation}}
\begin{document}
\title{Deeply quasi-bound state in single- and double-$\bar{K}$ nuclear clusters}

\author{S.~Marri}
\email{s.marri@ph.iut.ac.ir}
\affiliation{Department of Physics, Isfahan University of Technology, Isfahan 84156-83111, Iran}
\author{S. Z.~Kalantari}
\affiliation{Department of Physics, Isfahan University of Technology, Isfahan 84156-83111, Iran}
\author{J.~Esmaili}
\affiliation{Department of Physics, Faculty of Basic Sciences, Shahrekord University, Shahrekord, 115, Iran}
\date{\today}

\begin{abstract}
New calculations of the quasi-bound state positions in $K^{-}K^{-}pp$ kaonic nuclear cluster are 
performed using non-relativistic four-body Faddeev-type equations in AGS form. The corresponding 
separable approximation for the integral kernels in the three- and four-body kaonic clusters is 
obtained by using the Hilbert-Schmidt expansion procedure. Different phenomenological models of 
$\bar{K}N-\pi\Sigma$ potentials with one- and two-pole structure of $\Lambda$(1405) resonance 
and separable potential models for $\bar{K}$-$\bar{K}$ and nucleon-nucleon interactions, are used. 
The dependence of the resulting four-body binding energy on models of $\bar{K}N-\pi\Sigma$ interaction 
is investigated. We obtained the binding energy of the $K^{-}K^{-}pp$ quasi-bound state $\sim$ 80-94 
MeV with the phenomenological $\bar{K}N$ potentials. The width is about $\sim$ 5-8 MeV for the 
two-pole models of the interaction, while the one-pole potentials give $\sim$ 24-31 MeV width.
\end{abstract}
\pacs{
}
\maketitle
%%%%%%%%%%%%%%%%%%%%%%%%%%%%%%%%%%%%%%%%%%%%%%%%%%%%%%%%%%%%%%%%%%%%%%%%%%%%%%%%%%%%%%%%%%%%%%%%%%%
\section{Introduction}
\label{intro}
%%%%%%%%%%%%%%%%%%%%%%%%%%%%%%%%%%%%%%%%%%%%%%%%%%%%%%%%%%%%%%%%%%%%%%%%%%%%%%%%%%%%%%%%%%%%%%%%%%%
Over the past two decades, much attention has been placed on the studying of antikaon-nucleon and -nucleus interaction 
and the formation of dense $\bar{K}$ nuclear clusters~\cite{akaishi,yamazaki1,dote1}. The $K^{-}pp$ is the lightest 
possible kaonic nuclear bound system in which the proportion of strongly attractive $\bar{K}N$ ($I=0$) pairs to less 
attractive $\bar{K}N$ ($I=1$) pairs is maximized. Many theoretical works mostly focusing on the lightest kaonic system 
have been performed~\cite{dote1,dote2,shev1,shev2,revai,ikeda1,ikeda2,ikeda3,maeda}. All calculations confirm the 
existence of quasi-bound state in the $K^{-}pp$ system, but the values of the binding energy and width vary over a 
fairly wide range.

From the experimental point of view, this issue has also attracted considerable attention. The first experimental 
evidence concerning $K^{-}pp$ was observed in the stopped $K^{-}$ on $\mathrm{^{6,7}{Li}}$ and $\mathrm{^{12}{C}}$ 
targets~\cite{angello} by FINUDA collaboration at DA$\mathrm{\Phi}$NE. An exclusive analysis of the 
$p+p\rightarrow{X}+K^{+},X\rightarrow{p}+\Lambda$ experiment at Saclay for the $pp$ reaction at 2.85 
GeV~\cite{yamazaki2} indicated a large peak both in the $\Lambda{p}$ invariant-mass and $K^{+}$ missing-mass 
spectra, which had been predicted in the theoretical works~\cite{yamazaki3,yamazaki4}. Only when the object 
$X$ is a dense bound state of $K^{-}pp$ system, a peak comparable to the free emission of the $\Lambda^{\star}$ 
would be observed. The $K^{-}pp$ quasi-bound state has been further explored at J-PARC through 
$d(\pi^{+},K^{+})K^{-}pp$ and $K^{-}+\mathrm{^{3}{He}}$ reactions by E27~\cite{nagae2} and E15~\cite{hiraiwa2} 
experiments, respectively. However, so far the experimental studies on binding and width of $K^{-}pp$ neither 
agree with theoretical predictions nor their results are in accordance with each other. 

The $\bar{K}\bar{K}N$ system by quantum numbers $J^{\pi}=\frac{1}{2}^{+}$ and $I=\frac{1}{2}$ is 
also a possible three-body kaonic system. The $\bar{K}\bar{K}N$ quasi-bound state has been studied 
by Shevchenko and Haidenbauer using phenomenological and chiraly motivated potentials for the 
$\bar{K}N-\pi\Sigma$ interaction combined with Faddeev AGS equations~\cite{shev5}. This has also 
been investigated by Kanada-En'yo and Jido using a Gaussian expansion method~\cite{kanada}. In 
both studies, a quasi-bound state was found in the $\bar{K}\bar{K}N$ three-body system just a 
few MeV below the $\Lambda(1405)+\bar{K}$ threshold energy and it was shown that the repulsive 
interaction in $\bar{K}\bar{K}$ with $I=1$ makes the $\bar{K}\bar{K}N$ system loosely bound with 
moderate binding energy. The natural question which arises now is what will happen if we add an 
antikaon to the $K^{-}pp$ or a proton to the $K^{-}K^{-}p$ three-body system. In 2004, Akaishi 
and Yamazaki investigated the simplest double-$\bar{K}$ nuclear cluster, $K^{-}K^{-}pp$, using a 
phenomenological interaction based on the G-matrix method~\cite{yamaz}. They have shown that this 
system is deeply bound with binding energy of 117 MeV and could be considered as an important 
doorway toward multi-$\Lambda^{\star}$ nuclei. The existence of such a deeply kaonic nuclear state 
is important for studying high-density $\bar{K}$-nuclear systems~\cite{yamaz}, kaon condensation 
issue and neutron strars~\cite{kaplan}.

In this paper, we performed the nonrelativistic Faddeev-type Alt-Grassberger-Sandhas (AGS) calculations 
for the $K^{-}pp$ and $K^{-}K^{-}p$ three-body systems as well as for $K^{-}K^{-}pp$ four-body system. 
To convert the few-body Faddeev AGS equations to a manageable set of equations, we have to introduce 
the separable representation of the few-body amplitudes and the driving terms, which will be necessary 
to find the pole position of kaonic nuclear systems~\cite{grass,nadro}. Our few-body Faddeev calculations 
is based on the quasi-particle method. Using this method, we can find a separable representation for 
the subamplitudes in (3+1) and (2+2) partitions and one can reduce the three- and four-body problem 
to an effective quasi-particle two-body one, where one of the components appears as a quasi-particle. 
For making a separable representation of these subsystem amplitudes, one can use the energy dependent 
pole expansion (EDPE)~\cite{sofia} or the Hilbert-Schmidt expansion~\cite{nadro}. For this purpose, we 
will apply the Hilbert-Schmidt expansion.

The dependence of the three- and four-body kaonic clusters pole positions on two-body interactions is 
investigated. Several models of $\bar{K}N-\pi\Sigma$ interactions, which are derived phenomenologicaly, 
are used~\cite{shev3,shev4}. The potentials reproduce experimental data on elastic and inelastic $K^{-}p$ 
cross-sections and kaonic hydrogen atom. The $\bar{K}N-\pi\Sigma$ potentials are also constructed to 
produce a one- or two-pole structure of $\Lambda(1405)$ resonance. The double-$\bar{K}$ clusters 
contain the repulsive $\bar{K}\bar{K}$ interaction. Thus, the question which arises now is how much 
this interaction is important in double kaonic systems under study. In our few-body calculations, 
we used a phenomenological potential for the repulsive $\bar{K}\bar{K}$ interaction and the parameters 
of the potential obtained in such a way to reproduce the scattering length of the lattice QCD calculations~\cite{bean}.

The paper is organized as follows: in sect. \ref{formal}, we describe the framework of the present 
calculation and a brief description of Faddeev equations in the AGS form for three- and four-body kaonic 
nuclear systems is presented. Sect. \ref{put} is devoted to introducing the two-body inputs of the 
calculations. In sect. \ref{result}, we present our results of the three- and four-body calculations 
and the conclusions are presented in sect. \ref{conc}.
%%%%%%%%%%%%%%%%%%%%%%%%%%%%%%%%%%%%%%%%%%%%%%%%%%%%%%%%%%%%%%%%%%%%%%%%%%%%%%%%%%%%%%%%%%%%%%%%%%%
\section{Formalism}
\label{formal}
%%%%%%%%%%%%%%%%%%%%%%%%%%%%%%%%%%%%%%%%%%%%%%%%%%%%%%%%%%%%%%%%%%%%%%%%%%%%%%%%%%%%%%%%%%%%%%%%%%%
\subsection{Three-body equations}
The $\bar{K}NN$ and $\bar{K}\bar{K}N$ systems are coupled to $\pi\Sigma{N}$ and $\pi\bar{K}\Sigma$ 
channels, respectively. The three-body Faddeev-type AGS equations~\cite{grass,alt} for these systems are
\begin{equation}
U_{ij}^{\alpha\beta}=(1-\delta_{ij})\delta_{\alpha\beta}G_{0}^{-1}\sum_{k=1}^{3}
\sum_{\gamma=1}^{3}(1-\delta_{ik})T_{k}^{\alpha\gamma}G_{0}^{\gamma}U^{\gamma\beta}_{kj},
\end{equation}
where the operators $U^{\alpha\beta}_{ij}$ give the Faddeev amplitudes of the elastic and re-arrangement 
processes $i^{\alpha}+(j^{\alpha}k^{\alpha})\rightarrow{j}^{\beta}+(k^{\beta}i^{\beta})$ and the operators 
$T_{k}^{\alpha\gamma}$ are the two-body $T$-matrices embedded in the three-body space. The operator 
$G^{\alpha}_{0}$ is the free three-body Green's function; and the indices $i,j=1,2,3$ and $\alpha,\beta=1,2,3$ 
are used for describing the Faddeev partitions and particle channels, respectively~\cite{shev2}. Using the 
separable potentials for the two-body interactions
\begin{equation}
V_{i,I_{i}}^{\alpha\beta}(k,k')=g_{i,I_{i}}^{\alpha}(k_{\alpha})\lambda_{i,I_{i}}^{\alpha\beta}
g_{i,I_{i}}^{\beta}(k'_{\beta}),
\label{pot}
\end{equation}
will lead to a separable form of two-body $T$-matrices:
\begin{equation}
T_{i,I_{i}}^{\alpha\beta}(k,k';z)=g_{i,I_{i}}^{\alpha}(k_{\alpha} )\tau_{i,I_{i}}^{\alpha\beta}(z)
g_{i,I_{i}}^{\beta}(k'_{\beta}),
\label{tmat}
\end{equation}
where $I$ is a two-body isospin, $g^{\alpha}(k_{\alpha})$ are the usual form factors and 
$\tau_{i,I_{i}}^{\alpha\beta}$ being the usual two-body propagator. The three-body coupled 
channels Faddeev AGS equations for $\bar{K}NN-\pi\Sigma{N}$ and $\bar{K}\bar{K}N-\pi\bar{K}\Sigma$ 
systems are
\begin{equation}
\mathcal{K}_{ij,I_{i} I_{j}}^{\alpha\beta}=\delta_{\alpha\beta}
\mathcal{M}_{ij,I_{i}I_{j}}^{\alpha\beta}
+\sum_{k,I_{k};\gamma}\mathcal{M}_{ik,I_i I_k}^{\alpha}
\tau_{k,I_k}^{\alpha\gamma}
\mathcal{K}_{kj,I_k I_j}^{\gamma\beta},
\label{ags1}
\end{equation}
where the operator $\mathcal{K}_{ij,I_{i}I_{j}}^{\alpha\beta}$ is Faddeev transition amplitude 
between $\alpha$ and $\beta$ channels and the operator $\mathcal{M}_{ij,I_{i}I_{j}}^{\alpha\beta}$ 
is the effective potential, which are defined by
\begin{equation}
\mathcal{K}_{ij;I_{i} I_{j}}^{\alpha\beta}=\langle{g}^{\alpha}_{i,I_{i}}|G_{0}^{\alpha}U_{ij;I_{i}I_{j}}
G_{0}^{\beta}|{g}^{\alpha}_{j,I_{j}}\rangle,
\end{equation}
\begin{equation}
\mathcal{M}_{ij,I_{i}I_{j}}^{\alpha\beta}=\delta_{\alpha\beta}\mathcal{M}_{ij,I_{i}I_{j}}^{\alpha}=
\delta_{\alpha\beta}(1-\delta_{ij})\langle{g}^{\alpha}_{i,I_{i}}|G_{0}^{\alpha}|{g}^{\alpha}_{j,I_{j}}\rangle.
\end{equation}

The most important part of the quasi-particle approach is the separable representation of the off-shell 
Faddeev amplitudes in the two- and three-body systems. First of all, we will introduce the separable form 
of the three-body amplitudes and driving terms, for the $\bar{K}NN$ and $\bar{K}\bar{K}N$ systems by applying 
Hilbert-Schmidt expansion (HSE)~\cite{nadro}.
\begin{equation}
\mathcal{M}_{ij,I_i I_j}^{\alpha}(p,p',\epsilon)=-\sum^{N_{r}}_{n=1}\lambda_{n}(\epsilon)
u_{n;i,I_i}^{\alpha}(p,\epsilon)u_{n;j,I_j}^{\alpha}(p',\epsilon),
\label{ags}
\end{equation}
where $\lambda_{n}$ and the form factors $u_{n;i,I_i}^{\alpha}(p,\epsilon)$ are the eigenvalues 
and eigenfunctions of the kernel of equation (\ref{ags1}), respectively. The separable representation 
of the Faddeev AGS amplitudes is given by
\begin{equation}
\mathcal{K}_{ij,I_i I_j}^{\alpha\beta}(p,p',\epsilon)=\sum^{N_{r}}_{n=1}u_{n;i,I_i}^{\alpha}
(p,\epsilon)\zeta_{n}(\epsilon)u_{n;j,I_j}^{\beta}(p',\epsilon),
\label{ags2}
\end{equation}
and the functions $\zeta_{n}(\epsilon)$ obey the equation
\begin{equation}
\zeta_n(\epsilon)=\lambda_n(\epsilon)/(\lambda_n(\epsilon)-1).
\label{zeta}
\end{equation}

To search for a quasi-bound state, we should look for a solution of the homogeneous equations related to 
the form factors 
$u_{n;i,I_i}^{\alpha}(p,\epsilon)$
\begin{equation}
u_{n;i,I_i}^{\alpha}=\frac{1}{\lambda_n}\sum_{k=1}^{3}\sum_{\gamma=1}^{3}\sum_{I_k}
\mathcal{M}_{ik,I_i I_k}^{\alpha}\tau_{k,I_k}^{\alpha\gamma}u_{n;k,I_k}^{\gamma}.
\label{ags3}
\end{equation}

The AGS equation of (\ref{ags3}) is a Fredholm type integral equation. To solve the AGS 
equations for both $\bar{K}NN$ and $\bar{K}\bar{K}N$ systems, the operators involving 
two identical nucleons and kaons should be antisymmetric and symmetric, respectively. In 
$\bar{K}\bar{K}N$ system, the kaons are spinless particles, then all operators in 
isospin base involving two kaons, should be symmetric while in the case of $\bar{K}NN$, 
the spin component is antisymmetric (spin $s=0$). Thus, all operators in isospin base 
should be symmetric. To find a quasi-bound state in $\bar{K}NN$ and $\bar{K}\bar{K}N$ 
systems, one should convert the integral equations into algebraic form and then search 
for a complex energy at which the determinant of the kernel matrix is equal to zero.
%%%%%%%%%%%%%%%%%%%%%%%%%%%%%%%%%%%%%%%%%%%%%%%%%%%%%%%%%%%%%%%%%%%%%%%%%%%%%%%%%%%%%%%%%%%%%%%%%%%
\subsection{Four-body equations}
\label{fourbody}
\begin{figure*}[t]
%\vspace*{5.cm}
%\hspace*{-0.5cm}
\includegraphics[scale=0.7]{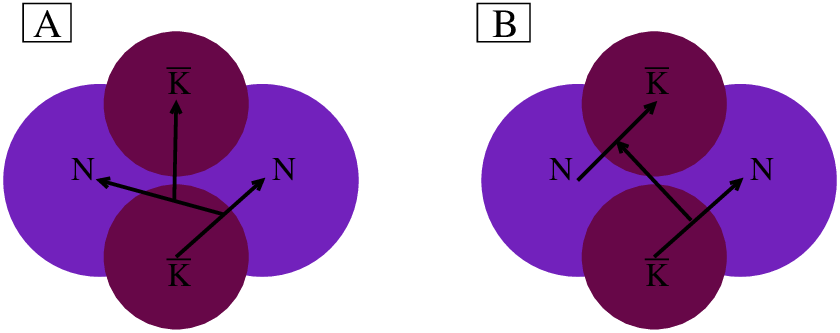}
%\vspace*{-4.5cm}
\caption{Two different rearrangement channels of $\bar{K}\bar{K}NN$ four-body system. (A) 
Channel 1, a two-body channel of (1+3) type (K-type); (B) channel 3, a two-body channel of 
(2+2) type (H-type). Antisymmetrization and symmetrization is to be made between two nucleons 
and between two kaons, respectively.}
\label{kkpp}
\end{figure*}

\begin{figure*}[t]
%\vspace*{2.5cm}
\includegraphics[scale=0.7]{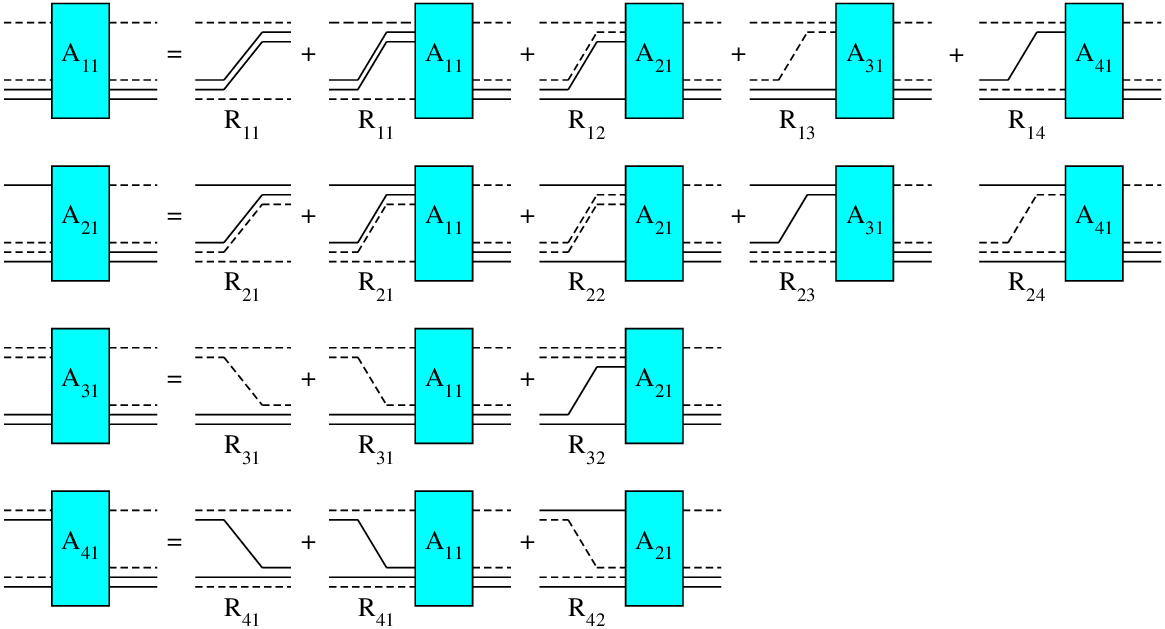}
%\vspace*{-2.5cm}
\caption{Diagrammatic representation of the equations (\ref{transpo}) for the Faddeev amplitudes 
$\mathcal{A}_{\alpha1}$ of the $\bar{K}\bar{K}NN$ system.}
\label{ags-equation}
\end{figure*}

%%%%%%%%%%%%%%%%%%%%%%%%%%%%%%%%%%%%%%%%%%%%%%%%%%%%%%%%%%%%%%%%%%%%%%%%%%%%%%%%%%%%%%%%%%%%%%%%%%%
In the present subsection, we briefly outline the formal aspects of the four-body Faddeev 
formalism applying to $\bar{K}\bar{K}NN$ system within the quasi-particle method in 
momentum space. Although a variety of methods for studying the four-body systems has been 
proposed in the literature, the Faddeev AGS method~\cite{grass,alt} and Faddeev-Yakubovsky 
approach~\cite{yak1,yak2} are more preferable to other methods. Using the separable approximation 
for the two-body potentials and for Faddeev amplitudes appearing in different K-type 
and H-type partitions of the four-body system, the both Faddeev approaches will produce the 
same set of effective two-body equations~\cite{grass,nadro,fonse}. Although the structure of 
the four-body equations are much more complicated compared to the three-body case, but currently, 
a practical formalism of four-particle theory has been extensively developed. Using properly 
symmetrized and antisymmetrized states with respect to identical kaons and nucleons, we will 
have the following four channels, corresponding to four possible two-quasiparticle partitions 
of $\bar{K}\bar{K}NN$ system. In fig. \ref{kkpp}, one K-type ($\bar{K}+[\bar{K}NN]$) and 
one H-type ($[\bar{K}N]+[\bar{K}N]$) configurations of the $\bar{K}\bar{K}NN$ four-body system 
are shown. The whole dynamics of $\bar{K}\bar{K}NN$ system is described in terms of the transition 
amplitudes $\mathcal{A}_{\alpha1}$ ($\alpha=$1,2,3 and 4) which connect the four quasi-two-body 
channels characterized by
\begin{equation}
\begin{split}
& \alpha=1:\bar{K}+(\bar{K}NN), \ \ \alpha=2:N+(\bar{K}\bar{K}N),\\
& \alpha=3:(\bar{K}N)+(\bar{K}N), \ \ \alpha=4:(\bar{K}\bar{K})+(NN),
\end{split}
\label{parti}
\end{equation}
with the initial channel $\alpha=1$. The essence of the calculation scheme is the solution of 
the bound state problem for the two- and three-body subsystems that is specified in the partitions 
(\ref{parti}). For $\alpha=1$ and 2 we dealt with interacting three-body systems. Using separable 
representations for the $NN$ and $\bar{K}N$ potentials, the corresponding scattering amplitudes 
can be expressed in terms of effective quasi-two-body amplitudes $\mathcal{K}^{\alpha\beta}_{ij,I_{i}I_{j}}$ 
which describe the scattering of a particle on a two-body cluster (quasi-particle). Due to the 
strong dominance of s-waves in $\bar{K}N$ and $NN$ interactions, we take into account only the 
s-wave part of the interactions in two, three and four-particle states. Then, we drop the 
index $l=0$ in all equations. Considering the identity of the kaons and the nucleons, the 
$\bar{K}\bar{K}NN$ problem is reduced to a set of $4\times4$ integral equations in one scalar 
variable. For the transition amplitudes $\mathcal{A}_{\alpha1}$ as a connector of channel 1 
to channels $\alpha =$1, 2, 3 and 4, we arrive at a coupled set of equations
\begin{equation}
\begin{split}
& \mathcal{A}^{sI,s'I'}_{\alpha\beta,nn'}(p,p',E)=\mathcal{R}^{sI,s'I'}_{\alpha\beta,nn'}(p,p',E) \\
& \hspace{2.5cm}+\sum^{3}_{\gamma=1}\sum_{n''s''I''}\int^{\infty}_{0}\mathcal{R}^{sI,s''I''}_{\alpha\gamma,nn''}(p,p'',E)  \\
& \hspace{2.5cm}\times \zeta_{\gamma,n''}(\epsilon_{\gamma})\mathcal{A}^{s''I'',s'I'}_{\gamma\beta,n''n'}(p'',p',E)d\vec{p''},
\end{split}
\label{transpo}
\end{equation}
where the operators $\mathcal{A}^{sI,s'I'}_{\alpha\beta,nn'}$ are the Faddeev amplitudes. The 
operators $\mathcal{R}^{sI,s'I'}_{\alpha\beta,nn'}$ are the effective potentials that are realized 
through particle exchange between the quasi-particles in channels $\alpha$ and $\beta$ and 
the arguments $\zeta_{\alpha,n}$ are the effective propagators, which are given in (\ref{zeta}). 
The AGS equations (\ref{transpo}) for the $\bar{K}\bar{K}NN$ system are schematically illustrated 
in fig. \ref{ags-equation}.
%%%%%%%%%%%%%%%%%%%%%%%%%%%%%%%%%%%%%%%%%%%%%%%%%%%%%%%%%%%
%%%%%%%%%%%%%%%%%%%%%%%%%%%%%%%%%%%%%%%%%%%%%%%%%%%%%%%%%%%
The effective potentials $\mathcal{R}^{sI,s'I'}_{\alpha\beta,nn'}(p,p',E)$ can be expressed in terms of the form 
factors $u^{sI}_{\alpha,n}$, which are generated by the separable representation of the sub-amplitudes appearing 
in the channels (\ref{parti})
\begin{eqnarray}
\begin{split}
& \mathcal{R}^{sI,s'I'}_{\alpha\beta,nn'}(p,p',E)=\frac{\Omega_{sI,s'I'}}{2}\int^{+1}_{-1}u^{sI}_{\alpha,n}
(\vec{q},\epsilon_{\alpha}-\frac{p^{2}}{2\mathcal{M}_{\alpha}}) \\ 
& \times\tau(z=E-\omega(p,p'))u^{s'I'}_{\beta,n'}(\vec{q'},\epsilon_{\beta}-\frac{p'^{2}}{2\mathcal{M}_{\beta}})
d(\hat{\vec{p}}\cdotp\hat{\vec{p}}').
\end{split}
\label{poten}
\end{eqnarray}
%%%%%%%%%%%%%%%%%%%%%%%%%%%%%%%%%%%%%%%%%%%%%%%%%%%%%%%%%%%
%%%%%%%%%%%%%%%%%%%%%%%%%%%%%%%%%%%%%%%%%%%%%%%%%%%%%%%%%%%

\begin{figure}[H]
%\vspace*{0.5cm} 
\hspace*{1.5cm}
\resizebox{0.8\textwidth}{!}{%
\includegraphics{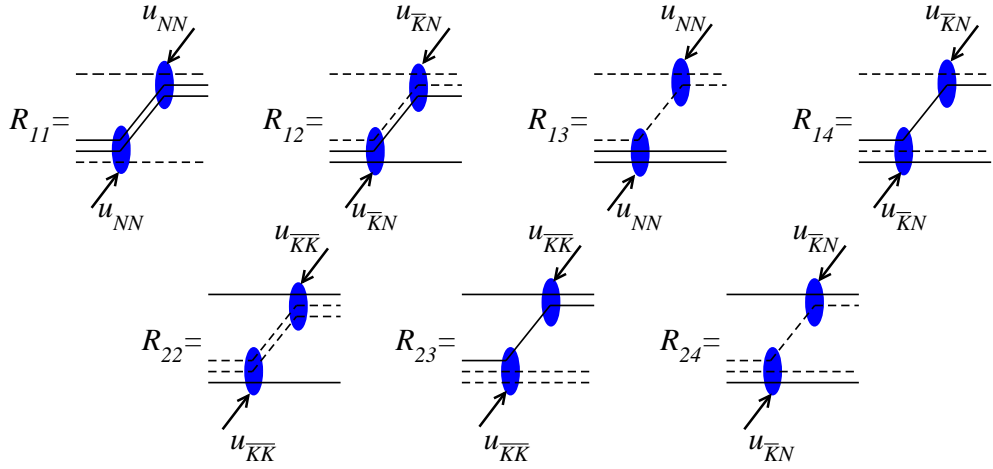}}
%\vspace*{-6.cm}
\caption{Diagrammatic representation of the potentials $\mathcal{R}_{\alpha\beta}$ in the separable 
approximation. The dashed line corresponds to the $\bar{K}$ and the solid lines corresponds to the 
nucleon. The symbols $u_{\alpha}$ define the initial and final states of the system.}
\label{zzz}
\end{figure}

Here, the symbols $\Omega_{sI,s'I'}$ are the spin-isospin Clebsch-Gordan coefficients, the 
argument $z=E-\frac{p^{2}}{2M_{\beta}}-\frac{p'^{2}}{2M_{\alpha}}-\frac{\vec{p}\cdot\vec{p}'}{m}$ 
is the energy of two-body quasi-particle, embedded in the four-body space and $\epsilon_{\alpha}$ 
is the total energy of the subsystem in channel $\alpha$. The effective potentials for 
$\bar{K}\bar{K}NN$ system are represented by the particle exchange diagrams in fig.~\ref{zzz}. 
The momenta $\vec{q}(\vec{p},\vec{p}')$ and $\vec{q}'(\vec{p},\vec{p}')$ are given in terms 
of $\vec{p}$ and $\vec{p'}$ by the following relations
\begin{equation}
\vec{q}=\vec{p}'+\frac{M_{\alpha}}{m}\vec{p}, \ \ \ \ \ \  \vec{q}'=\vec{p}+\frac{M_{\beta}}{m}\vec{p}',
\label{trans3}
\end{equation}
where $m$ is the exchanged particle or quasi-particle mass and the reduced masses $\mathcal{M}_{\alpha}$
and $M_{\alpha}$ in channel $\alpha$ are defined by
\begin{equation}
\begin{split}
& \mathcal{M}_{\alpha}=m^{\alpha}_{i}(m^{\alpha}_{j}+m^{\alpha}_{k}+m^{\alpha}_{l})/(m^{\alpha}_{i}+m^{\alpha}_{j}
+m^{\alpha}_{k}+m^{\alpha}_{l}), \\
& M_{\alpha}=m^{\alpha}_{j}(m^{\alpha}_{k}+m^{\alpha}_{l})/(m^{\alpha}_{j}+m^{\alpha}_{k}+m^{\alpha}_{l}),
\end{split}
\label{trans4}
\end{equation}
and in the case of H-type subsystems are given by
\begin{equation}
\begin{split}
& \mathcal{M}_{\alpha}=(m^{\alpha}_{i}+m^{\alpha}_{j})(m^{\alpha}_{k}+m^{\alpha}_{l})/(m^{\alpha}_{i}+m^{\alpha}_{j}
+m^{\alpha}_{k}+m^{\alpha}_{l}), \\
& M_{\alpha}=m^{\alpha}_{i}m^{\alpha}_{j}/(m^{\alpha}_{i}+m^{\alpha}_{j}).
\end{split}
\label{trans44}
\end{equation}
%%%%%%%%%%%%%%%%%%%%%%%%%%%%%%%%%%%%%%%%%%%%%%%%%%%%%%%%%%%%%%%%%%%%%
%%%%%%%%%%%%%%%%%%%%%%%%%%%%%%%%%%%%%%%%%%%%%%%%%%%%%%%%%%%%%%%%%%%%%
%%%%%%%%%%%%%%%%%%%%%%%%%%%%%%%%%%%%%%%%%%%%%%%%%
\begin{table*}[t]
\caption{Table of symbols that appear in the effective potentials $\mathcal{R}_{\alpha\beta}$. The symbols related to the 
driving terms $\mathcal{R}_{\beta\alpha}$, can be defined in a similar way.}
\centering
\begin{tabular}{cccc}
\hline\noalign{\smallskip}
$\mathcal{R}^{sI,s'I'}_{\alpha\beta,nn'}$ & $\vec{q}$ & $\vec{q}'$ & $\omega(p,p')$ \\
\noalign{\smallskip}\hline\hline\noalign{\smallskip}
\, $\mathcal{R}^{sI,s'I'}_{11,nn'}$ \, 
& \, $\vec{p}'+\frac{M_{\bar{K}}}{M_{\bar{K}}+2M_{N}}\vec{p}$ \, 
& \, $\vec{p}+\frac{M_{\bar{K}}}{M_{\bar{K}}+2M_{N}}\vec{p}'$ \, 
& \, $\frac{p^{2}(M_{\bar{K}}+2M_{N})}{2M_{\bar{K}}(M_{N}+M_{N})}+\frac{p'^{2}(M_{\bar{K}}+2M_{N})}{2M_{\bar{K}}(2M_{N})}
+\frac{\vec{p}\cdot\vec{p}'}{M_{N}+M_{N}}$  \, \\
\noalign{\smallskip}
\, $\mathcal{R}^{sI,s'I'}_{12,nn'}$ \, 
& \, $\vec{p}'+\frac{M_{N}}{M_{\bar{K}}+2M_{N}}\vec{p}$ \, 
& \, $\vec{p}+\frac{M_{\bar{K}}}{M_{N}+2M_{\bar{K}}}\vec{p}'$ \, 
& \, $\frac{p^{2}(M_{N}+2M_{\bar{K}})}{2M_{\bar{K}}(M_{N}+M_{\bar{K}})}+\frac{p'^{2}(M_{\bar{K}}
+2M_{N})}{2M_{N}(M_{N}+M_{\bar{K}})}+\frac{\vec{p}\cdot\vec{p}'}{M_{N}+M_{\bar{K}}}$  \, \\
\noalign{\smallskip}
\, $\mathcal{R}^{sI,s'I'}_{13,nn'}$ \, 
& \, $\vec{p}'+\frac{2M_{N}}{M_{\bar{K}}+2M_{N}}\vec{p}$ \, 
& \, $\vec{p}+\frac{1}{2}\vec{p}'$ \, 
& \, $\frac{p^{2}}{M_{\bar{K}}}+\frac{p'^{2}(M_{\bar{K}}+2M_{N})}{2M_{\bar{K}}(M_{N}+M_{N})}+\frac{\vec{p}
\cdot\vec{p}'}{M_{\bar{K}}}$  \, \\
\noalign{\smallskip}
\, $\mathcal{R}^{sI,s'I'}_{14,nn'}$ \, 
& \, $\vec{p}'+\frac{M_{N}+M_{\bar{K}}}{M_{\bar{K}}+2M_{N}}\vec{p}$ \, 
& \, $\vec{p}+\frac{M_{\bar{K}}}{M_{\bar{K}}+M_{N}}\vec{p}'$ \, 
& \, $\frac{p^{2}(M_{N}+M_{\bar{K}})}{2M_{\bar{K}}M_{N}}+\frac{p'^{2}(M_{\bar{K}}+2M_{N})}{2M_{N}(M_{N}+M_{\bar{K}})}
+\frac{\vec{p}\cdot\vec{p}'}{M_{N}}$  \, \\
\noalign{\smallskip}
\, $\mathcal{R}^{sI,s'I'}_{22,nn'}$ \, 
& \, $\vec{p}'+\frac{M_{N}}{2M_{\bar{K}}+M_{N}}\vec{p}$ \, 
& \, $\vec{p}+\frac{M_{N}}{2M_{\bar{K}}+M_{N}}\vec{p}'$ \, 
& \, $\frac{p^{2}(M_{N}+2M_{\bar{K}})}{2M_{N}(M_{\bar{K}}+M_{\bar{K}})}+\frac{p'^{2}(M_{N}+2M_{\bar{K}})}{2M_{N}(M_{\bar{K}}
+M_{\bar{K}})}+\frac{\vec{p}\cdot\vec{p}'}{2M_{\bar{K}}}$  \, \\
\noalign{\smallskip}
\, $\mathcal{R}^{sI,s'I'}_{23,nn'}$ \, 
& \, $\vec{p}'+\frac{2M_{\bar{K}}}{2M_{\bar{K}}+M_{N}}\vec{p}$ \, 
& \, $\vec{p}+\frac{1}{2}\vec{p}'$ \, 
& \, $\frac{p^{2}}{M_{N}}+\frac{p'^{2}(M_{N}+2M_{\bar{K}})}{2M_{N}(M_{\bar{K}}+M_{\bar{K}})}+\frac{\vec{p}\cdot\vec{p}'}{M_{N}}$  \, \\
\noalign{\smallskip}
\, $\mathcal{R}^{sI,s'I'}_{24,nn'}$ \, 
& \, $\vec{p}'+\frac{M_{N}+M_{\bar{K}}}{M_{N}+2M_{\bar{K}}}\vec{p}$ \, 
& \, $\vec{p}+\frac{M_{N}}{M_{\bar{K}}+M_{N}}\vec{p}'$ \, 
& \, $\frac{p^{2}(M_{N}+M_{\bar{K}})}{2M_{\bar{K}}M_{N}}+\frac{p'^{2}(M_{N}+2M_{\bar{K}})}{2M_{\bar{K}}(M_{N}+M_{\bar{K}})}
+\frac{\vec{p}\cdot\vec{p}'}{M_{\bar{K}}}$  \, \\
\noalign{\smallskip}\hline
\end{tabular}
\label{factors} 
\end{table*}
%%%%%%%%%%%%%%%%%%%%%%%%%%%%%%%%%%%%%%%%%%%%%%%%%
The corresponding assignments of the various symbols appearing in (\ref{poten}) for each 
effective potential are represented in table \ref{factors}. Before we proceed to solve the 
four-body equations, we also need to know the equations describing two independent pairs of 
interacting particles $(\bar{K}N)+(\bar{K}N)$ and $(\bar{K}\bar{K})+(NN)$ as input. In 
$(\bar{K}N)+(\bar{K}N)$ case, the corresponding equations read
\begin{equation}
\mathcal{Y}^{sI,s'I'}_{\bar{K}N,\bar{K}N}=\mathcal{W}^{sI,s'I'}_{\bar{K}N,\bar{K}N}+\mathcal{W}^{sI,s'I'}_{\bar{K}N,\bar{K}N}
\tau^{s'I'}_{NN}\mathcal{Y}^{s'I',s'I'}_{\bar{K}N,\bar{K}N},
\label{trans5}
\end{equation}
and the Faddeev equations for $(\bar{K}\bar{K})+(NN)$ system are defined by
\begin{equation}
\begin{split}
& \mathcal{Y}^{sI,s'I'}_{\bar{K}\bar{K},NN}=\mathcal{W}^{sI,s'I'}_{\bar{K}\bar{K},NN}+\mathcal{W}^{sI,s'I'}_{\bar{K}\bar{K},NN}
\tau^{s'I'}_{NN}\mathcal{Y}^{s'I',s'I'}_{NN,NN}, \\
& \mathcal{Y}^{s'I',s'I'}_{NN,NN}=  \mathcal{W}^{s'I',sI}_{NN,\bar{K}\bar{K}}
\tau^{sI}_{\bar{K}\bar{K}}\mathcal{Y}^{sI,s'I'}_{\bar{K}\bar{K},NN}.
\end{split}
\label{trans55}
\end{equation}

Here, $\mathcal{Y}^{sI,s'I'}_{i,j}$ are Faddeev amplitudes which describe two independent pairs of 
interacting particles and $\mathcal{W}^{sI,s'I'}_{i,j}$ are the effective potentials. Analogous 
to the treatment in the previous subsection, the separable form of the amplitudes can easily be found
\begin{equation}
\mathcal{Y}_{i,j}^{sI,s'I'}(p,p',\epsilon)=\sum^{N_{r}}_{n=1}u_{n;i}^{sI}(p,\epsilon)\zeta_{n}(\epsilon)
u_{n;j}^{s'I'}(p',\epsilon),
\label{trans6}
\end{equation}
where the functions $u_{n;i}^{sI}$ are the eigenfunctions of the kernel of eq. (\ref{trans5})
\begin{equation}
u_{n;i}^{sI}=\frac{1}{\lambda_n}\sum_{j=\bar{K}N,NN}
\mathcal{W}^{sI,s'I'}_{i,j}\tau^{s'I'}_{j}u_{n;j}^{s'I'}.
\label{trans7}
\end{equation}

Due to the strong coupling between $\bar{K}N$ and $\pi\Sigma$ channels, $\bar{K}\bar{K}NN$ four-body equations 
would be generalized to include the coupled channels $\bar{K}\bar{K}NN-\pi\bar{K}\Sigma{N}-\pi\pi\Sigma\Sigma$. 
There are seven different interactions in the lower-lying four-body channels, namely $\pi\pi$, $\pi\bar{K}$, 
$\pi\Sigma$, $\pi{N}$, $\Sigma\bar{K}$, $\Sigma\Sigma$, $\Sigma{N}$ and $\bar{K}N$. The $\pi\Sigma$ and $\bar{K}N$ 
interactions are included by using the coupled-channel model for $\bar{K}N-\pi\Sigma$ interaction. In practice, 
when we include the remaining interactions, the number of channels will increase rapidly and the treatment of the 
four-body equations turns out to be very complicated. Thus, the remaining interactions in the lower four-body 
channels are neglected for the system under consideration. This is necessary for faster convergence rate of the results. 

Before we proceed to solve the AGS equations (\ref{transpo}), we should antisymmetriz and symmetriz the basic 
amplitudes with respect to the exchange of nucleons and kaons, respectively. In practice to solve the four-body 
equations, it is necessary to convert the equations to a numerically manageable form by expanding (2+2) and (3+1) 
sub-amplitudes in eqs. (\ref{ags1}), (\ref{trans5}) and (\ref{trans55}) into separable series of finite rank 
$N_{r}$. We can use two different types of expansion. One is based on Hilbert-Schmidt expansion~\cite{nadro} 
method, and another one uses the energy dependent pole expansion (EDPE)~\cite{sofia}. In the present study, we 
have used Hilbert-Schmidt expansion (HSE) method.
%%%%%%%%%%%%%%%%%%%%%%%%%%%%%%%%%%%%%%%%%%%%%%%%%%%%%%%%%%%%%%%%%%%%%%%%%%%%%%%%%%%%%%%%%%%%%%%%%%%
\section{Two-body input}
\label{put}
%%%%%%%%%%%%%%%%%%%%%%%%%%%%%%%%%%%%%%%%%%%%%%%%%%%%%%%%%%%%%%%%%%%%%%%%%%%%%%%%%%%%%%%%%%%%%%%%%%%
In this section we shall begin with a survey on the two-body interactions, which are the inputs to our 
present study. The main $\bar{K}N-\pi\Sigma$ potential is constructed with orbital angular momentum $l=0$ 
since the interaction is dominated by $s$-wave $\Lambda$(1405) resonance. The $NN$, $\bar{K}\bar{K}$ 
and $\Sigma{N}-\Lambda{N}$ interactions were also taken in $l=0$ state and the remaining interactions are 
neglected in our calculations. All separable potentials in momentum representation have the form (\ref{pot}). 
%%%%%%%%%%%%%%%%%%%%%%%%%%%%%%%%%%%%%%%%%%%%%%%%%%%%%%%%%%%%%%%%%%%%%%%%%%
\subsection{$\bar{K}N-\pi\Sigma$ coupled-channel system}
\label{kn}
%%%%%%%%%%%%%%%%%%%%%%%%%%%%%%%%%%%%%%%%%%%%%%%%%%%%%%%%%%%%%%%%%%%%%%%%%%
The $\bar{K}N-\pi\Sigma$ interaction is the most important interaction of the three- and four-body kaonic 
nuclear systems. The $\bar{K}N$ interaction, is usually described either by pure phenomenological or by 
chirally motivated potentials. In our Faddeev calculations, we used four different effective potentials for 
the coupled-channel $\bar{K}N-\pi\Sigma$ interaction, having a one- and two-pole structure of the 
$\Lambda$(1405) resonance.
The potentials that we used here for the $\bar{K}N$ interaction are given in refs.~\cite{shev3,shev4}. 
The parameters of the coupled-channel $\bar{K}N-\pi\Sigma$ potential were fitted to reproduce all existing 
experimental data on the low-energy $K^{-}p$ scattering and kaonic hydrogen. The fitting was performed by 
using physical masses in $\bar{K}N$ and $\pi\Sigma$ channels with the inclusion of Coulomb interaction.
The parameters of $\bar{K}N-\pi\Sigma$ potential in ref.~\cite{shev3}, are adjusted to reproduce the most 
recent experimental results of the SIDDHARTA experiment~\cite{bazzi} and the one in ref.~\cite{shev4} 
reproduce the experimental results of the KEK experiment~\cite{kek1,kek2}. The form factors of the one-pole 
version and the $\bar{K}N$ channel of the two-pole version have a Yamaguchi form
\begin{equation}
g_{I}^{\alpha}(k^{\alpha})=\frac{1}{(k^{\alpha})^2+(\Lambda_I^{\alpha})^2},
\end{equation}
while a slightly more complicated form is used for the $\pi\Sigma$ channel
\begin{equation}
g_{I}^{\alpha}(k^{\alpha})=\frac{1}{(k^{\alpha})^2+(\Lambda_I^{\alpha})^2}
+\frac{s(\Lambda_I^{\alpha})^2}{((k^{\alpha})^2+(\Lambda_I^{\alpha})^2)^2}.
\end{equation}
%%%%%%%%%%%%%%%%%%%%%%%%%%%%%%%%%%%%%%%%%%%%%%%%%%%%%%%%%%%%%%%%%%%%%%%%%%
\subsection{$NN$ and $\Sigma{N}-\Lambda{N}$ interactions}
\label{nn}
%%%%%%%%%%%%%%%%%%%%%%%%%%%%%%%%%%%%%%%%%%%%%%%%%%%%%%%%%%%%%%%%%%%%%%%%%%
We also used one-term PEST potential from ref.~\cite{pest}, which is a separable approximation of the 
Paris model of $NN$ interaction. The strength parameter of PEST $\lambda_{NN}^{I}=-1$, and the 
form-factor is defined by
\be
g_{NN}^{I}(k)=\frac{1}{2\sqrt{\pi}}\sum_{i=1}^{6}\frac{c_{i;I}^{NN}}{(\beta_{i;I}^{NN})^{2}+k^{2}},
\ee
where a family of such $c_{i;I}^{NN}$ and $\beta_{i;I}^{NN}$ parameters are given in ref.~\cite{pest}. The 
on- and off-shell properties of the one-term PEST $NN$ potential is equivalent to the Paris potential up 
to $E_{lab}\sim 50$ MeV. It reproduces the triplet and singlet $NN$ scattering lengths, $a(^{3}S_{1})=-5.422$ fm 
and $a(^{1}S_{0})=17.534$ fm, respectively, as well as the deuteron binding energy $B.E_{\mathrm{deu}}=−2.2249$ MeV.

For the $s$-wave $\Sigma{N}-\Lambda{N}$ interaction, we follow the form given in ref.~\cite{tores},
\be
V_{\alpha\beta}^{I}(k,k')=-\frac{C_{\alpha\beta}^{I}}{2\pi^2}(\Lambda_\alpha\Lambda_\beta)^{3/2}
(\mu_\alpha\mu_\beta)^{-1/2}g_{\alpha}^{I}(k)g_{\beta}^{I}(k'),
\ee
where the symbols $C_{\alpha\beta}^{I}$ are the coupling constants summarized in table \ref{table3}, 
$\mu_{\alpha}$ is the reduced mass for the $\Sigma{N}$ and $\Lambda{N}$ system, the form factor 
$g_{\alpha}^{I}(k)$ is defined by  $g_{\alpha}^{I}(k)=1/(k^2+\Lambda_{\alpha}^{2})$, and the range 
parameters $\Lambda_{\alpha}$ are given by $\Lambda_{\Sigma{N}}=1.27\mathrm{fm}^{-1}$ and 
$\Lambda_{\Lambda{N}}=1.33\mathrm{fm}^{-1}$.
\begin{table}[ht]
\caption{Coupling constants of the $\Sigma{N}-\Lambda{N}$ interactions~\cite{tores}.} 
\centering 
\begin{tabular}{c c c c } 
\hline\hline 
$C_{\Sigma{N}-\Sigma{N}}^{I=1/2}$ & $C_{\Sigma{N}-\Lambda{N}}^{I=1/2}$ & $C_{\Lambda{N}-\Lambda{N}}^{I=1/2}$ 
& $C_{\Sigma{N}-\Sigma{N}}^{I=3/2}$  \\ [0.5ex] 
\hline 
0.83 & 0.56 & 0.49 & -0.29  \\ [1ex] 
\hline 
\end{tabular}
\label{table3} 
\end{table}
%%%%%%%%%%%%%%%%%%%%%%%%%%%%%%%%%%%%%%%%%%%%%%%%%%%%%%%%%%%%%%%%%%%%%%%%%%
\subsection{$\bar{K}\bar{K}$ interaction}
\label{kk}
%%%%%%%%%%%%%%%%%%%%%%%%%%%%%%%%%%%%%%%%%%%%%%%%%%%%%%%%%%%%%%%%%%%%%%%%%%
In contrast to the $\bar{K}N$ interaction, the amount of data on $\bar{K}$-$\bar{K}$ scattering 
($S=-2$) is scarce and the experimental situation is poorer than the above three two-body interactions. 
We introduce the effective interaction of the subsystem $\bar{K}\bar{K}$ with $I=1$, 
$V^{I=1}_{\bar{K}\bar{K}}$, in a Yamaguchi form
\begin{equation}
\begin{split}
& V^{I=1}_{\bar{K}\bar{K}}(k,k')=\lambda^{I=1}_{\bar{K}\bar{K}}g_{\bar{K}\bar{K}}(k)g_{\bar{K}\bar{K}}(k'), \\
& \ \ \ \ \ \ \ g_{\bar{K}\bar{K}}(k)=\frac{1}{k^{2}+\Lambda^{2}_{\bar{K}\bar{K}}}.
\end{split}
\end{equation}

During these calculations, we consider the $\bar{K}\bar{K}$ potentials with the parameters 
$\lambda^{I=1}_{\bar{K}\bar{K}}$ and $\Lambda_{\bar{K}\bar{K}}$, which reproduce the $K^{+}K^{+}$ scattering 
length, for which we used the result of lattice QCD calculation as $a_{K^{+}K^{+}}=−0.141$ 
fm~\cite{bean} as a guideline. The range parameter value 3.5 $\mathrm{fm}^{-1}$ is adopted for $\bar{K}\bar{K}$ interaction 
to represent the exchange of heavy mesons. 
%%%%%%%%%%%%%%%%%%%%%%%%%%%%%%%%%%%%%%%%%%%%%%%%%%%%%%%%%%%%%%%%%%%%%%%%%%%%%%%%%%%%%%%%%%%%%%%%%%%
\section{Results and discussion}
\label{result}
%%%%%%%%%%%%%%%%%%%%%%%%%%%%%%%%%%%%%%%%%%%%%%%%%%%%%%%%%%%%%%%%%%%%%%%%%%%%%%%%%%%%%%%%%%%%%%%%%%%
%%%%%%%%%%%%%%%%%%%%%%%%%%%%%%%%%%%%%%%%%%%%%%%%%%%%%%%%%%%%%%%%%%%%%
\begin{figure*} 
\begin{center}
\includegraphics[width=0.4\textwidth]{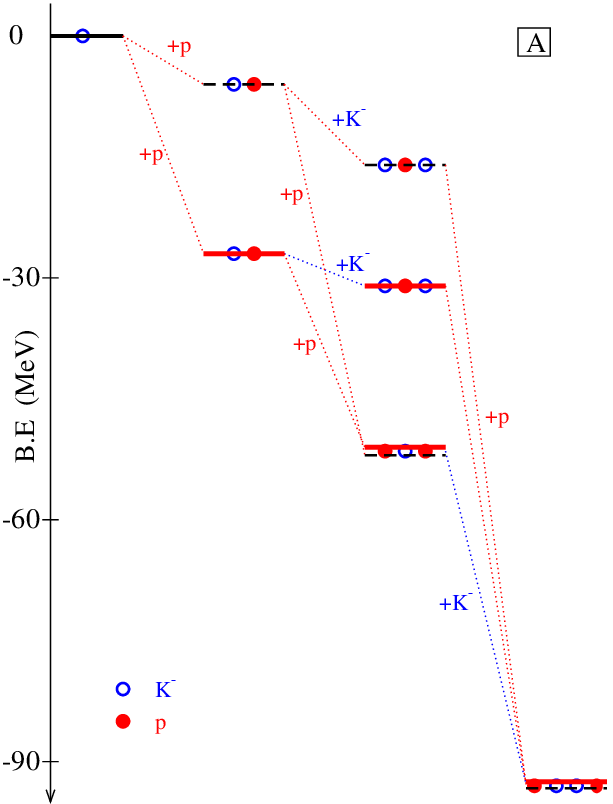}
\hspace*{1.cm}
\includegraphics[width=0.4\textwidth]{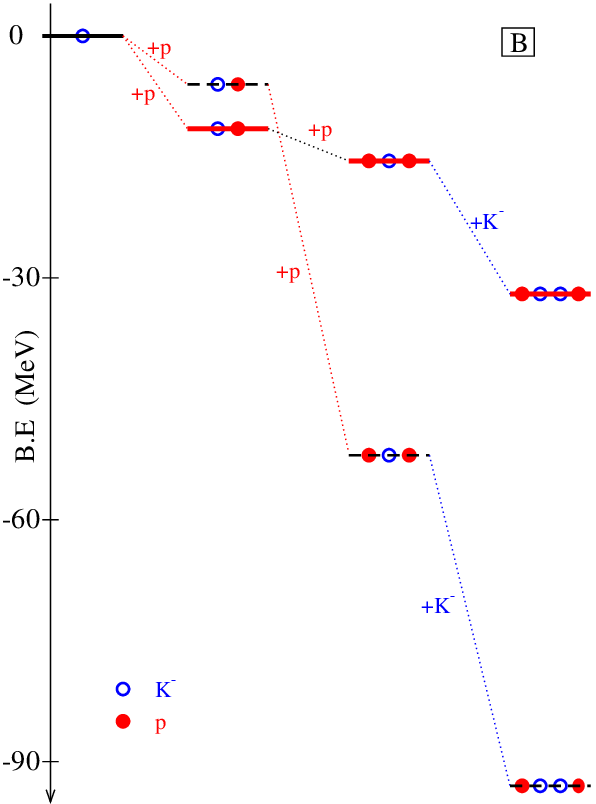} \\
\caption{(Color online) Global view of the calculated bound-state energies $B.E$ (in MeV) of the 
kaonic nuclear systems, $\bar{K}N$, $\bar{K}NN$, $\bar{K}\bar{K}N$ and $\bar{K}\bar{K}NN$. Diagram 
(A) shows a comparison between the present calculations using one-pole version of the 
SIDDHARTA potential $V^{SIDD,One-pole}_{\bar{K}N-\pi\Sigma}$ (black dashed lines) and the results 
by Maeda {\it et al.} using a simple one-channel real potential for the $\bar{K}N$ interaction 
~\cite{maeda} (red solid lines). Diagram (B) shows a comparison of the present calculations 
using the same potential in diagram (A) with the chiral-based results by Barnea {\it et al.}~\cite{gal} 
(red solid lines).}
\end{center}
\label{comp}
\end{figure*}
%%%%%%%%%%%%%%%%%%%%%%%%%%%%%%%%%%%%%%%%%%%%%%%%%%%%%%%%%%%%%%%%%%%%%
Solution of the Faddeev AGS equations corresponding to the bound and resonance states in the 
$(I,J^{\pi})=(\frac{1}{2},0^{-})$ and $(\frac{1}{2},\frac{1}{2}^{+})$ states of the $\bar{K}NN-\pi\Sigma{N}$ and 
$\bar{K}\bar{K}N-\bar{K}\pi\Sigma$ three-body systems, respectively, and $(I,J^{\pi})=(0,0^{+})$ state of 
$\bar{K}\bar{K}NN$ four-body system are found by applying search procedures described in sect. \ref{formal}. 
One- and two-pole version of the $\bar{K}N-\pi\Sigma$ interaction are considered and the dependence of the resulting 
few-body pole energy on the two-body $\bar{K}N-\pi\Sigma$ potentials is investigated. The $s$-wave (3+1) and (2+2) 
sub-amplitudes are obtained by using the Hilbert-Schmidt expansion (HSE) procedure for the integral kernels.

The calculated binding energies and the widths of the quasi-bound state of the $K^{-}pp$, $K^{-}K^{-}p$ and $K^{-}K^{-}pp$ 
systems for one- and two-pole of potentials are presented in table \ref{ta1}. The quasi-bound state position of the 
$K^{-}K^{-}pp$ system is obtained by keeping four terms ($N_{r}=4$) in the Hilbert-Schmidt expansion of the amplitudes 
(\ref{ags2}) and (\ref{trans6}), which will be suitable for practical calculations~\cite{kh}. It can be seen from table \ref{ta1} 
that our calculated binding energies are very close to the other results obtained in~\cite{revai} and~\cite{shev5} for 
$K^{-}pp$ and $K^{-}K^{-}p$ systems using the same $\bar{K}N-\pi\Sigma$ potentials within the coupled-channel Faddeev approach.

In the present calculations, the $\pi\bar{K}\Sigma{N}$ and $\pi\pi\Sigma\Sigma$ channels have not been 
included directly and one-channel Faddeev AGS equations are solved for the $\bar{K}\bar{K}NN$ 
system. We approximated the full coupled-channel one- and two-pole models of interaction by 
constructing the so-called exact optical $\bar{K}N-\pi\Sigma$ potential. The exact optical potential 
provides exactly the same elastic $\bar{K}N$ scattering amplitude as the coupled-channel model 
of interaction~\cite{shev4}. Thus, our coupled-channels four-body calculations with coupled-channel 
$\bar{K}N-\pi\Sigma$ interaction is equivalent to the one-channel four-body calculation using 
the so-called exact optical $\bar{K}N(-\pi\Sigma)$ potential. The decaying to the $\pi\bar{K}\Sigma{N}$ 
and $\pi\pi\Sigma\Sigma$ channels is taken into account through the imaginary part of the optical 
$\bar{K}N(-\pi\Sigma)$ potential. The binding energy and width of the deeply bound dibaryonic 
double-$\bar{K}$ system, $K^{-}K^{-}pp$, is calculated as a natural extension of $K^{-}pp$ and 
$K^{-}K^{-}p$ systems. The last row for each potential in table \ref{ta1} reports on the $K^{-}K^{-}pp$ 
quasi-bound state ($S=-2$) which has been highlighted as a possible doorway to kaon condensation 
in self-bound systems, given its large binding energy over 100 MeV predicted by Yamazaki 
{\it et al.}~\cite{yamaz}. The $K^{-}K^{-}pp$ system is tightly bound and has a larger binding 
energy than $K^{-}pp$, $B_{K^{-}K^{-}pp}\sim$80-94 MeV, and a width, $\Gamma=5-31$ MeV. In particular, it 
should be noted that the addition of one nucleon to the $K^{-}K^{-}p$ system gains $\sim$55-75 MeV, 
and the addition of one $\bar{K}$ to the $K^{-}pp$ system gains $\sim$35-40 MeV to the ground-state energy.
%%%%%%%%%%%%%%%%%%%%%%%%%%%%%%%%%%%%%%%%%%%%%%%%%%%%%%%%%%%%%%%%%%%%%
\begin{table}[H]
\caption{
The sensitivity of the pole position (in MeV), $z^{pole}_{X}$, of the quasi-bound states in $\bar{K}N$, 
$\bar{K}NN$, $\bar{K}\bar{K}N$ and $\bar{K}\bar{K}NN$ systems to the different models of $\bar{K}N-\pi\Sigma$ 
interaction. The real part of the pole position for each state is measured from the threshold of the 
corresponding kaonic system. $V^{One-pole}_{\bar{K}N-\pi\Sigma}$ and $V^{Two-pole}_{\bar{K}N-\pi\Sigma}$ 
standing for a one-pole and a two-pole structure of the $\Lambda$(1405) resonance.}
\centering
\begin{tabular}{ccc}
\hline\hline\noalign{\smallskip}\noalign{\smallskip}
 & $V^{One-pole}_{\bar{K}N-\pi\Sigma}$ & $V^{Two-pole}_{\bar{K}N-\pi\Sigma}$ \\
\noalign{\smallskip}\noalign{\smallskip}\hline
\noalign{\smallskip}\noalign{\smallskip}
With SIDD potential~\cite{shev3}: & & \\
\noalign{\smallskip}
$z^{pole}_{\bar{K}N}$  & $1428.1-i46.6$ & $1418.1-i56.9$ \\
\noalign{\smallskip}\noalign{\smallskip}
                    &                & $1382.0-i104.2$ \\
\noalign{\smallskip}\noalign{\smallskip}
$z^{pole}_{\bar{K}NN}$  & $-52.8-i31.5$ & $-48.5-i24.1$ \\
\noalign{\smallskip}\noalign{\smallskip}
$z^{pole}_{\bar{K}\bar{K}N}$  & $-17.8-i56.7$ & $-27.6-i41.2$ \\
\noalign{\smallskip}\noalign{\smallskip}
$z^{pole}_{\bar{K}\bar{K}NN}$  & $-92.7-i15.5$ & $-83.8-i4.0$ \\
\noalign{\smallskip}\noalign{\smallskip}
\hline
\noalign{\smallskip}\noalign{\smallskip}
With KEK potential~\cite{shev4}: & & \\
\noalign{\smallskip}
$z^{pole}_{\bar{K}N}$ & $1411.3-i35.8$ & $1410.8-i35.9$ \\
\noalign{\smallskip}\noalign{\smallskip}
                      &                & $1380.8-i104.8$ \\
\noalign{\smallskip}\noalign{\smallskip}
$z^{pole}_{\bar{K}NN}$ & $-44.5-i26.0$ & $-45.9-i19.5$ \\
\noalign{\smallskip}\noalign{\smallskip}
$z^{pole}_{\bar{K}\bar{K}N}$ & $-27.5-i38.1$ & $-26.7-i30.6$ \\
\noalign{\smallskip}\noalign{\smallskip}
$z^{pole}_{\bar{K}\bar{K}NN}$ & $-83.1-i12.4$ & $-81.5-i2.3$ \\
\noalign{\smallskip}\noalign{\smallskip}
\hline\hline
\end{tabular}
\label{ta1} 
\end{table}
%%%%%%%%%%%%%%%%%%%%%%%%%%%%%%%%%%%%%%%%%%%%%%%%%%%%%%%%%%%%%%%%%%%%%
In order to investigate the importance of repulsion between two kaons in double-$\bar{K}$ 
systems, we looked at the dependence of $K^{-}K^{-}p$ and $K^{-}K^{-}pp$ binding energies 
on the repulsive $K^{-}K^{-}$ interaction. In table \ref{ta2}, the $K^{-}K^{-}p$ and 
$K^{-}K^{-}pp$ binding energies for different representative sets of $\bar{K}N-\pi\Sigma$ 
potentials are obtained when the repulsive $V^{I=1}_{\bar{K}\bar{K}}$ is taken to be zero. 
It can be seen from the tables \ref{ta1} and \ref{ta2} that while the presence and absence of 
the repulsive $K^{-}K^{-}$ interaction can change the binding energy of the $K^{-}K^{-}p$ 
system about 5-10 MeV, the variation of the binding energy in the case of the $K^{-}K^{-}pp$ 
system is very small for all $\bar{K}N$ interaction models. Therefore, in contrast to 
$K^{-}K^{-}p$ system, the s-wave $K^{-}K^{-}$ interaction, which is used in the present 
calculation, plays a minor role in the $K^{-}K^{-}pp$ binding energy. Most likely, it is caused 
by the relative weakness of the $K^{-}K^{-}$ interaction as compared to $\bar{K}N$ from the 
viewpoint of a deep quasi-bound in the latter system ($B.E_{\bar{K}N}\sim 6-25$ MeV).
%%%%%%%%%%%%%%%%%%%%%%%%%%%%%%%%%%%%%%%%%%%%%%%%%%%%%%%%%%%%%%%%%%%%%
\begin{table}[H]
\caption{Pole positions, $z^{pole}_{X}$, of the quasi-bound states in $\bar{K}\bar{K}N$ and 
$\bar{K}\bar{K}NN$. In these calculations, the $\bar{K}\bar{K}$ interaction is turned off 
($\tau^{I=1}_{\bar{K}\bar{K}}=0$). The real part of the pole position for each state is measured 
from the threshold of the corresponding system. Comparing the present results with those in table \ref{ta1} 
shows that the binding energy of the $\bar{K}\bar{K}N$ system exhibits more sensitivity to the 
repulsive $\bar{K}\bar{K}$ interaction than the binding energy of the $\bar{K}\bar{K}NN$ four-body 
system. }
\centering
\begin{tabular}{ccc}
\hline\hline
\noalign{\smallskip}\noalign{\smallskip}\noalign{\smallskip}
$\bar{K}N$ interaction & $z^{pole}_{\bar{K}\bar{K}N}$ (MeV) & $z^{pole}_{\bar{K}\bar{K}NN}$  \\
\noalign{\smallskip}\noalign{\smallskip}\noalign{\smallskip}\hline
\noalign{\smallskip}\noalign{\smallskip}\noalign{\smallskip}
$V^{SIDD,One-pole}_{\bar{K}N-\pi\Sigma}$~\cite{shev3} & $-27.9-i55.1$ & $-93.7-i15.3$ \\
\noalign{\smallskip}\noalign{\smallskip}\noalign{\smallskip}
$V^{KEK,One-pole}_{\bar{K}N-\pi\Sigma}$~\cite{shev4} & $-33.2-i37.9$ & $-84.6-i12.1$ \\
\noalign{\smallskip}\noalign{\smallskip}\noalign{\smallskip}
$V^{SIDD,Two-pole}_{\bar{K}N-\pi\Sigma}$~\cite{shev3} & $-32.0-i35.9$ & $-84.2-i3.9$ \\
\noalign{\smallskip}\noalign{\smallskip}\noalign{\smallskip}
$V^{KEK,Two-pole}_{\bar{K}N-\pi\Sigma}$~\cite{shev4} & $-30.4-i27.2$ & $-81.8-i2.3$ \\
\noalign{\smallskip}\noalign{\smallskip}\noalign{\smallskip}\hline\hline
\end{tabular}
\label{ta2} 
\end{table}
%%%%%%%%%%%%%%%%%%%%%%%%%%%%%%%%%%%%%%%%%%%%%%%%%%%%%%%%%%%%%%%%%%%%%
Recently, some few-body calculations are performed on the lightest kaonic nuclei by the hyperspherical 
harmonics~\cite{gal} and the Faddeev method~\cite{maeda}. Barnea {\it et al.}~\cite{gal} made a 
hyperspherical harmonics calculation for four-body $K^{-}K^{-}pp$ nuclear quasi-bound state using an 
energy dependent chiral interaction model for $\bar{K}N$ interaction. In this calculation, a quasi-bound 
state with $I=0$ and $J^{\pi}=0^{+}$, was found with a binding energy about 32 MeV and a width of 80 MeV 
below the threshold energy of the $\bar{K}\bar{K}NN$ state. However, their results were criticized in 
ref.~\cite{revai}. A similar conclusion was also drawn by Maeda {\it et al.}~\cite{maeda} using a simple 
one-channel real potential for the $\bar{K}N$ interaction combined with the Faddeev-Yakubowsky method. 
The obtained binding energy for the $\bar{K}\bar{K}NN$ was about 93 MeV below threshold energy. Our results 
for binding energy values of the $K^{-}pp$, $K^{-}K^{-}p$ and $K^{-}K^{-}pp$ quasi-bound state are compared 
with other theoretical results in fig.~\ref{comp}. The results obtained in Faddeev calculation using the 
one-pole version of the $\bar{K}N-\pi\Sigma$ potential $V^{SIDD,One-pole}_{\bar{K}N-\pi\Sigma}$ are shown 
together with Faddeev results in~\cite{maeda} (Diagram A) and variational results~\cite{gal} (Diagram B). 
It is seen from fig.~\ref{comp} (Diagram B) that the energy dependent chiral $\bar{K}N$ potential leads to 
a more shallow quasi-bound state than the phenomenological one in three- and four-body systems. This is due 
to the energy dependence of the chiral potential. The comparison of our results for $\bar{K}\bar{K}NN$ 
obtained for PEST $NN$ interaction and the coupled-channel $\bar{K}N-\pi\Sigma$ interaction with standard 
binding energies calculated in ref.~\cite{maeda} within the Faddeev-Yakubowsky method for rank-two $NN$ 
interaction and one-channel real $\bar{K}N$ interaction shows that they are in the same order of magnitude 
(Diagram A). Although the present results for the quasi-bound states in the $K^{-}pp$ and $K^{-}K^{-}pp$ 
systems and the binding energies reported by Maeda {\it et al.} are in the same range, but this agreement 
seems to be rather accidental. We want to emphasize that in fact it is difficult to compare our results with 
those in Maeda {\it et al.}~\cite{maeda}. Firstly, as already said in the sect. \ref{put}, the potential 
that we used here for the $\bar{K}N-\pi\Sigma$ interaction are adjusted to reproduce the experimental data 
on kaonic hydrogen and low-energy $K^{-}p$ scattering, but Maeda {\it et al.}~\cite{maeda} fixed the two-body 
energy arbitrarily to define the parameters of the $\bar{K}N$ potential. Secondly, in ref.~\cite{maeda} the 
$\pi\Sigma$ channel has not been included. The $\pi\Sigma$ channel plays an important dynamical role in forming 
the three- and four-body quasi-bound state. Thus, it is expected that the inclusion of $\pi\Sigma$ channel will 
have a serious effect on their obtained binding energy for $K^{-}pp$ and $K^{-}K^{-}pp$ quasi-bound states.
%%%%%%%%%%%%%%%%%%%%%%%%%%%%%%%%%%%%%%%%%%%%%%%%%%%%%%%%%%%%%%%%%%%%%%%%%%%%%%%%%%%%%%%%%%%%%%%%%%%
\section{Conclusion}
\label{conc}
%%%%%%%%%%%%%%%%%%%%%%%%%%%%%%%%%%%%%%%%%%%%%%%%%%%%%%%%%%%%%%%%%%%%%%%%%%%%%%%%%%%%%%%%%%%%%%%%%%%
In the present paper, non-relativistic four-body Faddeev equations have been applied 
to study the $\bar{K}\bar{K}NN$ system. The calculation scheme, which formally allows 
an exact solution, is based on the separable approximation of the appropriate integral 
kernels. We employed HSE method to reduce the problem to a set of single-variable 
integral equations. To investigate the dependence of the resulting four-body binding 
energy on models of $\bar{K}N-\pi\Sigma$ interaction, different versions of 
$\bar{K}N-\pi\Sigma$ potentials, which produce the one- or two-pole structure of 
$\Lambda$(1405) resonance, were used. We have also found that $K^{-}$-$K^{-}$ repulsion 
inside $K^{-}K^{-}pp$ in contrast to $K^{-}K^{-}p$ system, gives only a small effect 
on its binding energy and width, which does not alter the dense nature of this double-
$\bar{K}$ cluster. The calculations yielded binding energies $B_{K^{-}pp}\sim$ 45-53, 
$B_{K^{-}K^{-}p}\sim$ 17-28 and $B_{K^{-}K^{-}pp}\sim$ 80-94 MeV for $K^{-}pp$, $K^{-}K^{-}p$ 
and $K^{-}K^{-}pp$ systems, respectively. The obtained widths for these systems are 
$\Gamma_{K^{-}pp}\sim$ 40-62, $\Gamma_{K^{-}K^{-}p}\sim$ 60-110 and $\Gamma_{K^{-}K^{-}pp}\sim$ 
5-31 MeV. The calculations suggest that few-body double-$\bar{K}$ nuclei, such as $K^{-}K^{-}pp$, 
as well as single-$\bar{K}$ nuclei, are tightly bound systems with large binding energies. 
The results of the one-channel AGS calculations of $K^{-}K^{-}pp$ show that, if the 
difference between the two sets of the $K^{-}K^{-}pp$ binding energies corresponding to 
the one- and two-pole versions of the coupled-channel $\bar{K}N-\pi\Sigma$ potential is 
much more than theoretical uncertainties, then it would be possible to favor one version 
of $\bar{K}N-\pi\Sigma$ potential by comparing with an experimental result. Similar 
calculations could be performed for the chiraly motivated $\bar{K}N$ input potential, too. 
The quasi-bound states resulting from the energy-dependent potentials happen to be shallower 
because of the weaker $\bar{K}N$ attraction for lower energies than the energy independent 
potentials under consideration in this work.
%%%%%%%%%%%%%%%%%%%%%%%%%%%%%%%%%%%%%%%%%%%%%%%%%%%%%%%%%% 

The authors are thankful to A. Fix for his helpful discussions. The authors also gratefully acknowledge 
the Sheikh Bahaei National High Performance Computing Center (SBNHPCC) for providing computing 
facilities and time. SBNHPCC is supported by the scientific and technological department of presidential 
office and Isfahan University of Technology (IUT).
%%%%%%%%%%%%%%%%%%%%%%%%%%%%%%%%%%%%%%%%%%%%%%%%%%%%%%%%%%%%%%%%%%%%%%%%%%%%%%%%%%%%%%%%%%%%%%%%%%%
%\bigskip
%\newpage
%%%%%%%%%%%%%%%%%%%%%%%%%%%%%%%%%%%%%%%%%%%%%%%%%%%%%%%%%%%

\end{document}